\documentclass{aa}                             
\usepackage{graphicx}                              
\usepackage{natbib}
\usepackage{subfigure}
\usepackage{ulem}
\include{definitions}

      \def\new#1 {{\bf #1 }}
      \def\cut#1 {\sout{#1} }

\def\meth {$\mathrm{CH_3OH}$} 
\def\kms {$\mathrm{km\,s^{-1}}$} 
\def\HII {H{\sc ii}} 
\def\AMM {$\mathrm{NH_3}$} 
\def\percc {$\mathrm{cm^{-3}}$} 

\def\WAT {$\mathrm{H_2O}$} 
\def\HII{H{\sc ii}}
\def\simgreat{\mathbin{\lower 3pt\hbox
{$\rlap{\raise 5pt\hbox{$\char'076$}}\mathchar"7218$}}}
\def\simless{\mathbin{\lower 3pt\hbox {$\rlap{\raise 5pt\hbox{$\char'074$}}\mathchar"7218$}}} 
\begin{document}

\title{High mass star formation in the infrared dark cloud
G11.11$-$0.12}
\author{T. Pillai, F. Wyrowski, K. M. Menten, \&\ E.Kr\"ugel, }
\institute{Max-Planck-Institut f\"ur Radioastronomie,
Auf dem H\"ugel 69, D-53121 Bonn, Germany}
\offprints{T. Pillai, thushara@mpifr-bonn.mpg.de}

\date{\today}

\titlerunning{high-mass star formation in G11.11}
\authorrunning{Pillai et al.}

\abstract{ We report detection of moderate to high-mass star formation
in an infrared dark cloud (G11.11-0.12) where we discovered class II
methanol and water maser emissions at 6.7 GHz and 22.2 GHz,
respectively.  We also observed the object in ammonia inversion
transitions. Strong emission from the (3,3) line indicates a hot (~60
K) compact component associated with the maser emission. The line
width of the hot component (4 $\mathrm{km\,s^{-1}}$), as well as the
methanol maser detection, are indicative of high mass star formation.
To further constrain the physical parameters of the source, we derived
the spectral energy distribution (SED) of the dust continuum by
analysing data from the 2MASS survey, HIRAS, MSX, the Spitzer Space
Telescope, and interferometric 3mm observations.  The SED was modelled
in a radiative transfer program: {\it a)} the stellar luminosity
equals $\sim$1200 L$_\odot$ corresponding to a ZAMS star of 8
M$_\odot$; {\it b)} the bulk of the envelope has a temperature of 19
K; {\it c)} the mass of the remnant protostellar cloud in an area
$8\times 10^{17}$ cm or 15$''$ across amounts to 500\,M$_\odot$, if
assuming standard dust of the diffuse medium, and to about 60
M$_\odot$, should the grains be fluffy and have ice mantles; {\it d)}
the corresponding visual extinction towards the star, $A_{\rm V}$, is
a few hundred magnitudes.  The near IR data can be explained by
scattering from tenuous material above a hypothetical disk.  The class
II methanol maser lines are spread out in velocity over 11 km/s.  To
explain the kinematics of the masing spots, we propose that they are
located in a Kepler disk at a distance of about 250 AU.  The dust
temperatures there are around 150 K, high enough to evaporate
methanol--containing ice mantles.

}

\maketitle

\section{Introduction}

Infrared dark clouds (IRDCs) are cold, dense molecular clouds seen silhouetted
against the bright diffuse mid-infrared emission of the Galactic plane.  They
were discovered during mid-infrared imaging surveys with the Infrared Space
Observatory (ISO) \citep{perault1996:iso} and the Mid-course Space Experiment
(MSX\footnote{This research made use of data products from the Midcourse Space
Experiment.  Processing of the data was funded by the Ballistic Missile Defense
Organization with additional support from the NASA Office of Space Science.  This research also made use of the NASA/ IPAC Infrared Science Archive, which is
operated by the Jet Propulsion Laboratory, California Institute of Technology,
under contract with the National Aeronautics and Space Administration.}) 
\citep{egan1998:irdc}.  
The mm/submm observations revealed that typical IRDCs have gas
densities of $n > 10^{6} ~ {\rm cm}^{-3}$, temperatures of $T < 20 ~ {\rm K}$,
and sizes of 1-10~pc \citep{carey1998:irdc}. To determine the mass distribution
toward the IRDCs, Carey et al.\ (2000) obtained images of the 450 and 850 $\mu$m
emission using SCUBA. All observed IRDCs contain 1--4 bright sub-millimeter
emission peaks ($>$ 1 Jy/beam at 850 $\mu$m) surrounded by an envelope of
emission, which matches the morphology of the IRDC in mid-infrared extinction.

\citet{carey2000:irdc} suggest that IRDCs might be sites of early phases of
high-mass star formation, however studies of their star formation content are still
rare. There are, indeed two cases where detailed studies were done towards
the IRDCs, although the original object identification was not based on the
extinction in the MIR.  Based on the submm dust continuum and line observations,
\citet{sandell2000:ngc6334} conclude that NGC6334 I(N) is a high-mass Class 0 object.
Similarly, \citet{garay2002:msfr} suggest that IRAS16272 is a dense massive core
in a very early evolutionary stage, distinguished by being luminous without any
associated Ultracompact H{\sc ii} regions (UCH{\sc ii}R).  Toward the IRDC G79.3+0.3 P1, \citet{redman2003:g79} 
reported associated low and intermediate mass YSOs.  Also, \citet{teyssier2002:irdc} found young embedded stars associated with the IRDCs they studied,
but do not give further details on them.  Observations of IRDC G11.11-0.12 are
discussed by  \citet{johnstone2003:g11} and, although they noticed signs of star
formation associated with their SCUBA source P1, they did not find evidence of
massive star formation. Identifying the earliest
stages in the formation of massive stars is currently a crucial step in our
understanding of massive star formation, so we started a project to
search for methanol masers toward cold massive cores.


Class II methanol (\meth) masers (CIIMMs) are almost always found in
high-mass star-forming regions (HMSFRs).  
It turned out that not all of the CIIMM are associated with UCH{\sc
ii}R \citep{walsh1998:meth_radio}.  Based on high spatial-resolution 
radio continuum and 6.67~GHz methanol spectral-line data in
364 sources towards IRAS-selected regions,\citet{walsh1998:meth_radio}
show that the methanol maser is most likely present before an
observable UCH{\sc ii}R is formed around a massive star and is quickly
destroyed as the UCH{\sc ii}R evolves. However, this is disputed by
\citet{phillips1998:meth_maser} who argue that the non detection of
radio continuum emission associated with methanol masers is because of
the association of the maser sites with low/intermediate mass stars
which do not ionise their immediate surroundings. To date {\it no}
CIIMM\ has ever been detected toward a low-mass SFR, with the possible
exception of NGC 2024-FIR4 \citep{minier2003}, which, by the way, is
the {\it only} known CIIMM in the whole Orion Giant Molecular Cloud
(GMC) complex.  \citet{walsh2001} in another study of 31 methanol
maser sites and 19 UCH{\sc ii}R in Mid-infrared (10.5 and 20 $\mu$m)
find that the observations are consistent with the maser emission
being powered by the MIR source.  \citet{walsh2001} also find strong
evidence of infall/outflow in different molecular
tracer. \citet{beuther2002:masers} find a mean separation of $\sim
0.19$~pc between the \meth\ and cm emission for there sample of 29
massive star forming regions. Subsequently, in a continuum mapping
reported by \citet{walsh2003:dust_maser} toward 71 6.7~GHz maser
fields, they find that most methanol masers are within 10$''$ of a
sub-mm peak. Furthermore, a recent 1.2~mm continuum survey in massive
star forming regions with methanol maser sites and/or radio continuum
emission, reported by \citet{hill2005:1.2mm} shows that these tracers
of massive star formation are associated with heavily embedded
protostars.  These arguments taken together with the non detection of
radio continuum make a convincing case of CIIMMs signposting the
earliest stages of high mass star formation.

Here we report the discovery of CIIMM emission toward the IRDC
G11.11$-$0.12 and discuss the properties of the protostar heavily
embedded in the IRDC detected with SPITZER in the MIR.  In
\S\ref{sec:observations}, we describe our observation in various
molecular tracers with the Effelsberg 100-m telescope, Australian
Telescope Compact Array, Very Large Array and the
Berkeley-Illinois-Maryland-Association interferometer. In
\S\ref{sec:results} we present the evidences for active star formation
in the IRDC and model the ammonia emission. The discussion on the
Spectral Energy Distribution (SED) of the embedded source and
modelling in spherical symmetry follows through in
\S\ref{sec:discussion}.  The observed properties of the CIIMM are
entirely comparable with the properties of numerous other such sources
detected in high-mass star forming regions. Therefore, its presence
makes G11.11$-$0.12 , an IRDC in which active high-mass star formation
has been unequivocally established.

\section{Observations\label{sec:observations}}

\subsection{CH$_3$OH maser}

We observed the class II methanol maser at 6.7 GHz ($5_1-6_0\,$A$^+$ transition)
toward the submm peak IRDC G11.11 P1 ($\alpha_{\rm J2000}=18:10:28.402$,
$\delta_{\rm J2000}=-19:22:29.00$) with the Effelsberg 100-m telescope
\footnote{Based on observations with the 100-m telescope of the MPIfR
(Max-Planck-Institut für Radioastronomie) at Effelsberg} on 15 June 2003.  The
frontend was the facility 5 cm receiver tuned to 6668.518 MHz.  We used the
8192~channel auto-correlator with 2 subunits of 10~MHz bandwidth to be able to
reach a spectral resolution of $\approx 0.25$~km$\,$s$^{-1}$.  The beam width of
122$\arcsec$ was determined from drift scans over compact continuum sources,
which also served to check the pointing.  The observations were repeated with
the 100-m telescope on 5 August 2004.

On 11 January 2004, we followed up the observations of the 6.7 GHz Class II
\meth\ maser with the Australian Telescope Compact Array $(ATCA)$ in the
FULL\_4\_1024-128 correlator configuration, employing the snapshot mode with a
series of 6 short integration cuts of 5 minutes.  Since ATCA can observe
two frequencies simultaneously, we chose the 8.64~GHz continuum band covering a
bandwidth of 128~MHz frequencies. The 1 $\sigma$ sensitivity achieved is
$\sim 0.2$~mJy/beam.  Phase calibration was done using B1817-254.  The primary
beam of ATCA at this frequency is $\sim 8~'$, and the effective HPBW of the
synthesized beam equals 2.03$''$ $\times$ 6.24$''$.  The spectral resolution was
0.2 km/s, and the 1 $\sigma$ sensitivity 0.5~Jy.  The ATCA spectrum is
essentially identical to the 2003 Effelsberg spectrum.

\subsection{Ammonia observation}

The maser detection prompted us to search for the \AMM\ (3,3) and (4,4)
transitions with the Effelsberg 100-m telescope in April 2004.  With the AK 8192
backend, we were able to observe the (1,1), (2,2), (3,3), and (4,4) transitions
in both polarisations simultaneously using the K-band receiver.  With 8 subunits
of 10~MHz bandwidth, the resulting spectral resolution was $\approx
0.2$~km$\,$s$^{-1}$ after smoothing the data to improve the signal-to-noise
ratio.  The beam at the \AMM\ frequencies was 40$''$.  The observations were
performed in the frequency switching mode.  Pointing was checked at hourly
intervals by continuum scans on G10.62. We estimate, the pointing to be accurate within 6$''$, with the pointing scans used for the absolute calibration.

\subsection{VLA \WAT\ maser observation}

The 22.2~GHz \WAT\ maser observations were done with the Very Large Array (VLA)
on 24 August 2004 in its D configuration and in 2 polarisations with a spectral
resolution of 0.33 km/s, a synthesized HPBW of $\sim 0.4''$, and a primary beam
of $\sim 120''$.  The standard interferometer mode was used with a total
integration time of 20 minutes on source, split into sessions.  Phase calibration
was done using J1733-130, and the flux calibrator was 3C286.

\subsection{BIMA observation}

The 3mm observations were carried out from September 1999 to July 2000 with the
Berkeley-Illinois-Maryland-Association interferometer (BIMA) in its B, C, and D
configuration sampling spatial structures from 3$''$ to 60$''$.  Continuum
images were obtained with a combination of 86 and 93~GHz tracks centred with a
beam size of 8.3$''$ $\times$ 3.9$''$ reaching an rms of 1.6~mJy.  The total
bandwidth of both sidebands was $\sim900$~MHz.

\section{Results\label{sec:results}}

\subsection{CH$_3$OH maser}

\begin{figure}
\centering
 \includegraphics[height=\linewidth,angle=-90]{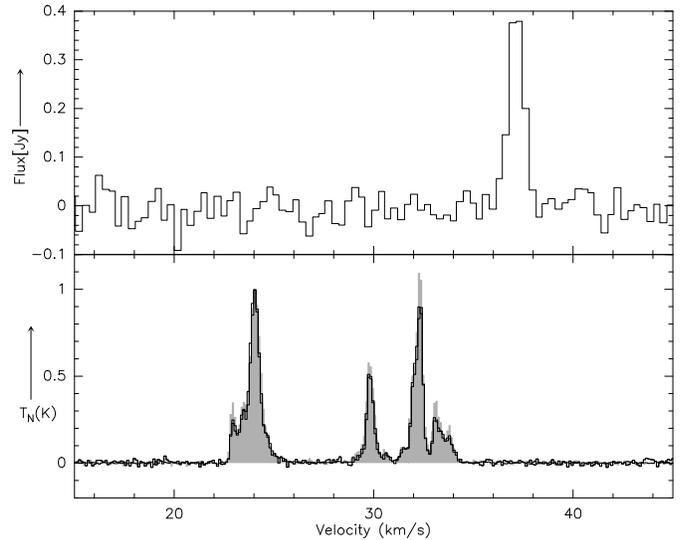}
 \caption{Lower panel: Effelsberg 100-m Spectra of the \meth\ emission at
 6.7~GHz.  The filled spectrum corresponds to June 2003 and the solid lines to
 August 2004 observations.  The blue shifted component is normalised to unity in
 both cases.  Upper panel: 22.2~GHz water maser spectra obtained with the
 VLA.}
\label{fig:meth_spec}
\end{figure}

The solid line in Fig.\ref{fig:meth_spec} displays the 6.7 GHz
methanol maser spectrum toward G11.11P1 detected in August 2004 with
the Effelsberg 100-m telescope, while the filled region shows the June
2003 data. As the Effelsberg 100-m absolute calibration has an
uncertainty of $\sim 20 \%$, the observations taken with two different
instruments and at different times have been normalised with respect
to one of the features, so that absolute calibration is no longer an
issue.  The maser features cover a velocity range of about 11\,km/s,
with the peaks at 32\,km/s in June 2003 and at 24\,km/s in August
2004.  Thus there is clear evidence of a variation in the relative
line intensities of about 10~\%.  From mapping the source, we find
that the masing spots originate at the position P1 from an area that
is point--like with respect to the Effelsberg beam.  There is no
evidence of another maser source.  The follow--up observations
obtained six months later with ATCA reveal no obvious changes in the
spectrum and also indicate that the maser originates from a single
core. The peak flux of the brightest maser feature from the ATCA
observations is 22~Jy.

\label{fig:1}

\label{fig:2}
\begin{figure*}
\centering \includegraphics[width=\linewidth,angle=0]{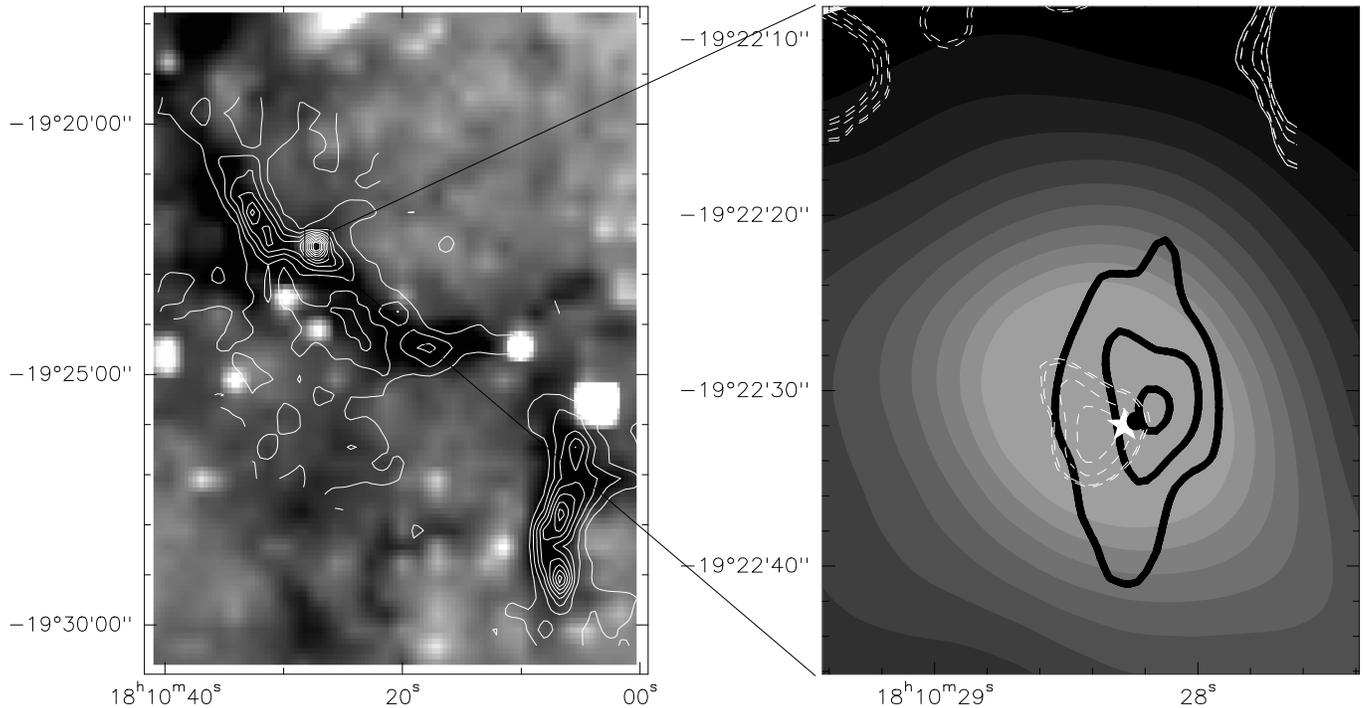}
 \caption{\textit{Left:} The $8 ~ \rm \mu m$ image of G11.11 with
 SCUBA $850 ~ \rm \mu m$ (Carey et al.\ 2000) overlay \textit{Right:}
 BIMA 3mm continuum image towards P1 in contours, with contour
 level$\sim$1.6~mJy (-3,3,5,7) and 2MASS $K_s$ band in dashed contours
 on the $850~ \rm \mu m$ image.  The star denotes the \WAT\ maser and
the filled circle the \meth\ maser position.}

\label{fig:g11image}
\end{figure*} 

The position of the integrated maser emission is plotted in Fig.\ref{fig:g11image}.
 There, the left panel
shows the MSX map of the filamentary dark cloud overlaid on the SCUBA submm map
from \citet{carey2000:irdc}, and the right panel is a blow--up of the SCUBA image
around P1, together with contours of 2MASS~$K_{s}$\footnote{The atlas image
obtained is part of the Two Micron All Sky Survey (2MASS), a joint project of
the University of Massachusetts and the Infrared Processing and Analysis
Center/California Institute of Technology, funded by the NASA and the NSF} band
and the 3mm continuum emission.  To estimate the positional accuracy of the maser
observations, we split the phase calibrator observations into two sets and cross-calibrate the second set with a self--calibrated first set, which results in a
position accuracy of (0.05$''$,-0.15$''$). This is a lower limit, but since our
source is less than $10\degr$ away from the calibrator, baseline errors are
not likely to add to the error.  To cross-check this, we reduced other
observations towards an already known maser source and found a positional
accuracy of (0.105$''$,-0.75$''$).  ATCA is an east west array,
so the beamwidth in declination is greater by a factor cosec(DEC); and for
those sources with declination north of -24 deg, complete u-v coverage is
unobtainable.

\begin{figure}
\centering
\includegraphics[height=8cm,angle=-90]{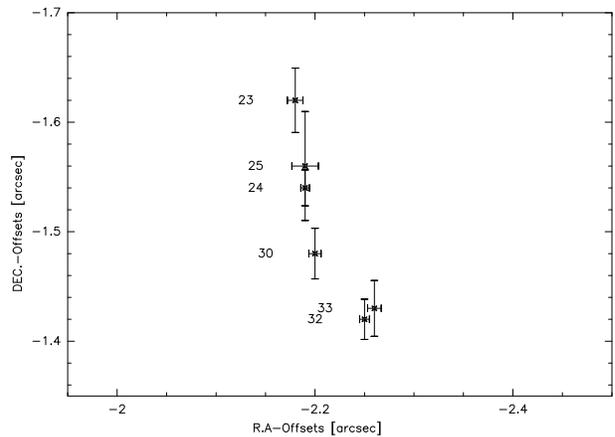}
 \caption{Velocity structure of the \meth\ maser emission. Each data point
 gives the position of the model fits to the individual velocity channels.  The
 corresponding $V_{LSR}$ in \kms is also shown. }
\label{fig:vel_structure}
\end{figure}

 When the sub-spots are analysed in position--velocity plots, they fall nicely
 along an arc, as predicted for a disk (\citealt{minier1998}, \citealt{norris1998}).  
There is also a linear velocity trend.  This is shown in Fig.\ref{fig:vel_structure} where
 a point source model was fitted to individual velocity channels.  The one sigma
 position uncertainty is proportional to the SNR \citep{reid1988}; hence,
 with a SNR~15-30 for the different velocity features, we reach a sub-arcsecond
 accuracy.  The relatively weak peak at 34 km/s was not included due to its
 large positional uncertainty.

\label{fig:3}

\subsection{3mm continuum emission}

The BIMA image shows a compact, slightly resolved 3mm continuum
source that coincides with the SCUBA sub-mm peak shown in
Fig.\ref{fig:g11image}.  A 2-D Gaussian fit to the emission yields a
source size of $0.25\times0.12$~pc with a PA $3^\circ$.  The peak flux
of $\sim 12$~mJy corresponds to a brightness temperature of $\sim
28$~mK.  Our cm continuum observations with ATCA at 8.4~GHz, along
with searches in Galaxy-wide cm continuum surveys, reveal no
detectable free-free emission.  In order to check whether one expects
to find free-free emission at 3mm, we extrapolated the ATCA upper
limit to that at 3mm, assuming optically thick emission.  We
estimated a mm flux of 25.3~mJy relative to the ATCA upper limit of
$\sim 0.2$~mJy.  Thus the contribution from free-free emission cannot
be ruled out.  For a typical value of 10000~K for the ionisation
temperature from an optically thick ionised gas cloud, we get an
emission measure $EM>4\times10^{10}$~pc~$\rm{cm}^{-3}$ at 90~GHz , a
very high value that is more suitably assigned to Hypercompact \HII\
regions \citep{kurtz2002}.  Such a region with typical sizes $<
0.05$~pc would have been point-like with respect to the BIMA beam,
while we found that the source is partially resolved. Thus, the 3mm
emission is most likely due to dust.

\subsection{\WAT\ maser}

The position of the \WAT\ maser appears offset from the \meth\ maser
position, as well as from the BIMA 3mm continuum peak, within the
respective positional errors.  This might indicate that, although both
\meth\ and \WAT\ masers occur in warm and dense environments, they may
not spatially coincide due to different excitation mechanisms
\citep{beuther2002:masers}.

Water masers can be explained by collisional pumping with H$_2$
molecules in shocks associated with outflows \citep{elitzur1989}.
Alternatively, the pumping can occur by accretion shocks in disks
\citep{garay1999}. The maser is weak ($\sim 0.3~$Jy) with a feature
offset of $\sim 7$~\kms from the systemic velocity as shown in the
upper panel of Fig.\ref{fig:meth_spec}. This agrees well with the
picture of water masers being variable and having large velocity
spread.

The maser spot is not spatially resolved over its velocity structure.
We performed a model fit to the individual velocity channels over the
single feature.  Unlike the \meth\ masers, the \WAT\ maser spots do
not show any systematic gradient.  Since the velocity structure is
also found to be spread on a much higher angular scale than the \meth\
maser, it may be tracing the outflow.  Higher angular resolution is
needed to reveal structures that could be either a shock front of the
high-velocity outflow very close to the central object or part of the
accretion disk itself.  We also serendipitously discovered another
water maser that was offset by more than 1$'$.  The feature is
brighter and situated at the systemic velocity of the source.  The
absolute position of the maser peak is at $\alpha_{\rm
J2000}$=18:10:33.58, $\delta_{\rm J2000}$=-19:22:50.3.

\subsection{Ammonia towards G11.11P1 \label{subsec:amm}}

\begin{figure}[h]
\centering \includegraphics[height=\linewidth,angle=0]{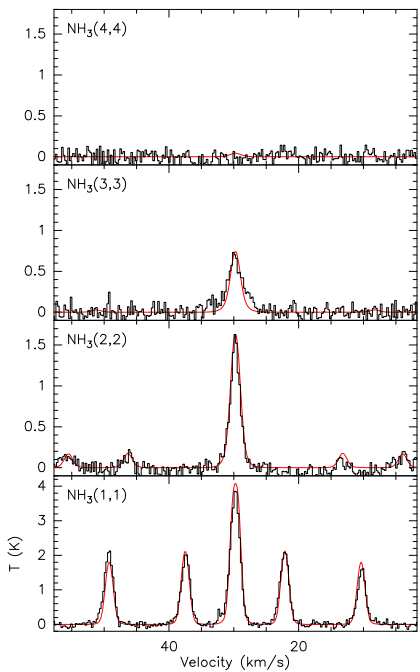}
  \caption{\textit{top to bottom}: Spectra of the NH$_3$ $(J,K)=(1,1), (2,2),
(3,3)$, and (4,4) transitions.  All spectra were taken towards the submm peak
position given in the text. The fit discussed in \S\ref{subsec:amm} is 
overlaid in red.}
\label{fig:nh3_spec}
\end{figure}

Figure~\ref{fig:nh3_spec} shows the ammonia lines observed with the 100-m
telescope toward G11.11P1, and the line parameters are given in
Table~\ref{tab:irdc_line_parameter}.  For the (4,4) line that was not
detected, the 1$\sigma$ r.m.s noise in the spectrum is given instead.

\begin{table}[h]
\caption{\sc Effelsberg 100-m line parameters }
\vspace{1em}
\begin{tabular}{lrccc}
\hline
\hline
 Transition & $v_{\rm LSR}$ & $T_{\rm MB}$ & 
     FWHM & \\
                 & (km$\,$s$^{-1}$)    & (K)  & (km$\,$s$^{-1}$) & \\
\hline
   \AMM\ 1--1 & 29.78(0.01) & 4.1 (0.4)  &  1.29(0.02)&\\
    \AMM\ 2--2 &  29.78(0.02) & 1.58(0.07)& 1.81 (0.06)&\\
   \AMM\ 3--3 &  29.77(0.07) & 0.64(0.05)& 3.22 (0.21)&\\
    \AMM\ 4--4 &   & (0.065) &  &\\
\hline
\end{tabular}
\label{tab:irdc_line_parameter}
\end{table}

\label{fig:4}
 The line widths from Gaussian fits when taking the hyperfine
 satellites for the (1,1) and (2,2) spectra into account increases
 with the excitation of the lines.  The (3,3) transition even shows,
 a broad wing besides the line core.  From a Boltzmann plot of the
 ammonia emission, we conclude that the ammonia emission cannot be
 explained with a single temperature. From just the \AMM\ (1,1) and
 (2,2) data, we got a rotational temperature of $\sim 14 ~ \rm K$.  The
 rotational temperature derived from the (2,2) and (3,3) lines was
 estimated to be $\sim 30~ \rm K$.

This evidence of two temperature components prompted us to fit the
spectrum with two components simultaneously using XCLASS (Schilke, P.,
private communication).  The program then yields the rotation
temperature, column density and source size for a given telescope size
assuming a Gaussian structure, and produces a synthetic spectrum in
agreement with observations.  Table 2 lists the fit results.  Thus the
broader component appears to be much hotter (60 K) and is very compact
(3.3$''$). Actually, the fit is not unique since there is degeneracy
between column density and the source size.  But a large source size
would mean a still higher luminosity, which would put the source in
the spectral type O9.  It is highly unlikely that an HII region around
such a star goes undetected.  On the other hand, a much smaller source
size with very high column density would mean extremely dense gas
with critical density high enough to excite the (4,4) level, which we
do not detect.

Additionally, we fit the ammonia spectra with a spherical
symmetric model \citep{wyrowski2003} using both a core component
with a temperature power law, consistent with internal heating from a
1200 L$_\odot$ embedded source, and a more extended, cooler (10~K)
lower density clump. For the core component, the luminosity, and the
outer and inner radii were taken from the dust continuum modelling, which will
be presented in \S\ref{subsec:mir}.  Assuming an ammonia abundance of
$10^{-8}$, the density in the core is $7 \times 10^5$~\percc, which is
consistent with the modelling in \S\ref{subsec:mir}, and is a factor 10
lower in the cooler clump with a size of 40\arcsec. For the core
component, a line width gradient was modelled with 4 km/s in the
inner 2 arcsec and then decreasing to the clump component with 1.2
km/s.  The increasing line widths/turbulence towards the centre of the
core is another indication of an embedded, massive young stellar
object that is stirring up its environment, as well as the remaining line wings
of the (3,3) line, which indicates an additional outflow component.

\begin{table}[h]
\caption{\sc Model fit results  }
\vspace{1em}
\begin{tabular}{lrcccc}
\hline
\hline
 Component &  $T_{\rm rot}$ & size  &  $N_{\rm NH_3}$&
        FWHM & \\
         & (K)    & ($''$)  &   ($10^{15} ~ \rm cm^{-2}$)&(km$\,$s$^{-1}$) & \\
\hline
   \AMM\ cold  & 15.4 & 23 &  1.8&  1.34&\\
    \AMM\ hot &  60 & 3.3 & 13.7 &4 &\\
   
\hline
\end{tabular}
\label{tab:model_fit}
\end{table}

\subsection{Archival Data}

We collected the data from 4 galactic sky survey archives, namely
2MASS, Spitzer, MSX, and IRAS.  The 2MASS point source catalog
contains the JHK fluxes for 3 point sources all within the 14$''$ of
the SCUBA beam.  The Spitzer MIR data collected from the Spitzer data
archive is part of the GLIMPSE Legacy survey in 4 of the Infrared
Array Camera (IRAC) bands from $3.6-8 \mu$m.  We found 3 point sources
in the $3.6\mu$m band, while in the higher wavelength bands we found a
single source that were coincident with the MSX $8\mu$m point source.
The FIR emission obtained from the high resolution IRAS data is not
point like and might suffer from background emission.  But IRAS point
sources generally do not appear circular, primarily because the
detector aperture were rectangular.  We adopted an flux uncertainty of
20\% in all cases as given by the respective Space centres.

\section{Discussion\label{sec:discussion}}

\subsection{\meth\ maser disk scenario}

Figure~\ref{fig:meth_spec} contains 3 major peaks that is very
reminiscent of a classical maser triplet, which is usually ascribed to
a maser amplification from a disk in Keplerian rotation viewed edge on
\citep{elmergreen1979}.

According to \citet{cesaroni1990} and \citet{ponomarve1994}, the
intensity ratio of the central to the red and blue peaks is determined
by the width of the ring in which the conditions for line
amplification are favourable, and the central peak vanishes when the
ring becomes very small. The red and blue components of our spectrum
are not equidistant to the central component that marks the systemic
velocity (see also Table~\ref{tab:model_fit} of \AMM\ velocities).
Such a distortion may be due to a tilt of the disk, in which case the
physical parameters (density, temperature, and velocity) along the line
of sight depend not only on the radial distance, but also on the
height above the disk.

In principle, equidistant triplets can also arise from an expanding
shell.  We favour the disk scenario for several reasons.  {\it a)}
Variations in the line ratios, as argued by \citet{cesaroni1990}, are
readily explained for disks by radiative interaction between masing
spots, whereas such an interaction would not be possible for an
expanding shell because of opposite velocities of the masing.  {\it
b)} All 3 major peaks contain sub-peaks, which we interpret as arising
from irregularities associated with the tilt or warp of the disk.
{\it c)} The near-infrared emission and the Spitzer data discussed in
\S 4.1.2 also seem to require a disk structure.

\subsection{Mass \& luminosity estimates}

The kinematic distance to G11.11 derived as 3.6~kpc using the rotation curve
of Clemens (1985) and the IAU standard values of $R_0 = 8.5$~kpc and
$V_0 = 220$~\kms by Carey et al.\ (1998) is 3.6~kpc.
\citet{carey2000:irdc} quote a gas mass of 67 M$_\odot$ from their dust
continuum observations at 850 micron.  The virial masses estimated from ammonia
observations range from $150-240~M_\odot$ for the hot and the cold components
within the beam (40$''$) of Effelsberg.  Similarly, by assuming the
fractional abundance of \AMM\ (relative to $H_2$) to be on the order of
$10^{-8}$, the \AMM\ column density translates to masses of $200-1600 M_\odot$.

The CIIMM need elevated temperatures for two reasons: first, high methanol
abundances are required to produce an adequate maser gain,. and second,
high temperatures are needed to produce pump photons.  To effectively
evaporate methanol off icy grain mantles, temperatures in excess of
100 K are needed (Sandford \& Allamandola 1993).  Recent models of
CIIMM pumping require gas temperatures $\sim 100$ K, somewhat higher
dust temperatures, and densities of about $10^7$ cm$^{-3}$.
Therefore, the compact \meth\ maser emission we observe must arise
from a hot, dense source.  We may adopt the spread of the maser
components, $0.25''$ or 900 AU, as the size of that source, which is
comparable to the sizes of numerous other CIIMM regions.  The
Stefan-Boltzmann law delivers with a luminosity of $830~L_\odot$
for a source of the corresponding area and $T = 100$ K . To fit all observed
ammonia lines simultaneously, we presented a model in \S\ref{subsec:amm}
that assumes an optically thin spherical dust cloud illuminated by a
central point source, which then results in a radial temperature distribution of $T
\propto r^{-0.4} $(Wilner et al.\ 1995). With a luminosity of $1200~L_\odot$,
we could explain the observed warm ammonia core embedded within a larger cold
clump. In the model, the temperature reaches a temperature of 240~K at the
inner radius, hence large enough to excite the methanol maser observed.

\subsection{SED\label{subsec:sed}}

\subsubsection{MIR and longer wavelengths \label{subsec:mir}}

\begin{figure}
\centering \includegraphics[height=\linewidth,angle=-90]{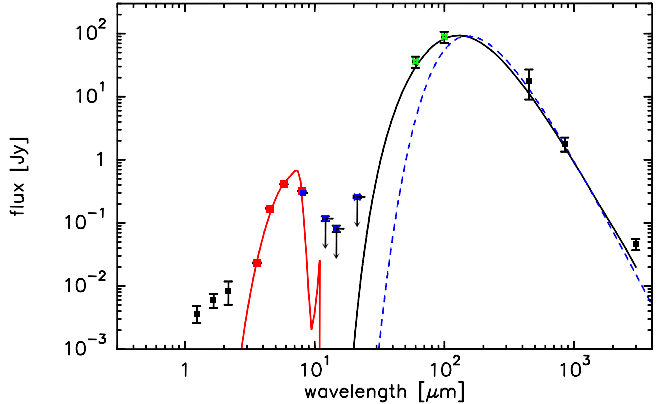}
 \caption{The Spectral Energy Distribution of G11.11-0.12 P1.
 Squares with errorbars indicate observations, where two almost
 coincide at 8 and 8.2$\mu$m. The upper limits from MSX corresponding
 to bands C (12.13$\mu$m ), D (14.65$\mu$m), and E (21.3$\mu$m) are
 shown. The solid curve represents the semi--spherical model described
 in the text. The solid line at FIR wavelengths corresponds to the SED
 of a central star of $L=1200$ L$_\odot$ and to the 15000~K surface
 temperature that is surrounded by an envelope. Possible second
 blackbody emitter is shown as solid curve at shorter
 wavelengths. The dashed curve represents a grey body fit of 19 K. The
 NIR fluxes is explained by scattering (see \S\ref{subsec:2mass})}

\label{fig:sed}
\end{figure}

The SED of G11.11 P1 shown in Fig.\ref{fig:sed} contains fluxes due to dust
emission at (i) 3mm (BIMA, 8$'' \times 4''$ beam), (ii) at 450 and $850\mu$m (JCMT), 
(iii) IRAS data, (iv )MSX upper limits at 12 and 21 $\mu$m, (v) an MSX detection at 8.2 $\mu$m,
(vi) four detections with the Spitzer Space Telescope (at 3.6, 4.5, 5.8 and
8.0$\mu$m), and (vii) three NIR data points from 2MASS.

Although for massive young stars, a time sequence from class 0 to class 3 has not
been established, a high--mass--star equivalent to a class 0 object should also
emit more than half a percent of its energy at submm wavelengths \citep{andre1993:class0}. 
 For G11.11, the luminosity at $\lambda > 300\mu$m amounts to almost
3\% of the total, so in this respect it could be considered an object equivalent
to a Class 0 source. Similar arguments of a Class 0 equivalent object in 
high-mass-star-forming region have also been proposed by \citet{motte2003} in
the submm protoclusters in W43 .  They derive a~submillimeter~to bolometric
luminosity ratio of 1.5-3\% in W43-MM1 and MM2.  

To estimate how the dust temperature and density vary within the envelope, we
tried to reproduce the continuum emission in a self-consistent way by
calculating the radiative transfer in spherical symmetry.  However, when
including the NIR fluxes, as well as the recently published data from the Spitzer
archive, the assumption of strict radial symmetry had to be abandoned.  We
sketch below a possible, or even likely, structure of the source; but to
confirmat its correctness we need information with higher spatial resolution, a difficult
requirement for such a distant source.  The model therefore remains rather
qualitative.

Nevertheless, the observations yield a number of important constraints.  {\it
a)} The total luminosity is determined by the IRAS fluxes. While the
100$\mu$m flux may contain a fair contribution by diffuse emission (from cloud
heating by the interstellar radiation field or scattered stellar sources), the
60$\mu$m flux indicating warm dust comes from a compact source with a luminosity
of $\sim$10$^3$ L$_\odot$.  {\it b)} As the source is only marginally resolved
at 3\,mm, its outer cloud radius $r_{\rm out}$ has to be smaller than $4\times
10^{17}$ cm.  {\it c)} The submm/mm fluxes lead to a total mass of several
hundred M$_\odot$ when adopting standard interstellar dust with an absorption
coefficient at 1\,mm of $\sim$$4\times 10^{-3}$ cm$^2$ per gram of interstellar
matter \citep{krugel2003}.  However, as this emission is optically thin, a
substantial part may not be directly linked to the star, but may come from the
background within the IRDC.  {\it d)} The visual extinction to the star depends
on how the dust is distributed.  It is about 400\,mag if all the mass is in a
sphere with constant density around the star, and even higher for a centrally condensed
envelope, and less than 400\,mag in case of an additional background cloud.  For
comparison,  we derive a column density $A_{\rm V} > 60$ mag from the peak of the ammonia emission in a 40$''$ beam.
This column density is probably two or three times larger,
depending on the compactness of the source. We then assumed conversion factors
[NH$_3$]/[H$_2$] = $10^{-8}$ and $N_{\rm H} = 2\times 10^{21}$ $A_{\rm V}$
cm$^{-2}$ \citep{motte1998:ophiuchus}.

The solid line in Fig.\ref{fig:sed}, which is compatible with the observations between
10$\mu$m and 3mm, comes from a spherical radiative transfer model (see Appendix
A) with a central star of $L=1200$ L$_\odot$ and 15000~K surface temperature
that is surrounded by a envelope whose inner and outer radii are
$10^{15}$ and $3.5\times10^{17}$ cm, respectively.  The exact number for $T_{\rm
eff}$ is irrelevant, because the stellar radiation gets immediately absorbed.  
The dust density in the envelope is constant ($\rho_{\rm dust} = 3
\times 10^{-20}$ g cm$^{-3}$) implying an optical extinction $A_{\rm V} \simeq
400$ mag.  It may, at most, be two times less, otherwise the model spectrum
would not comply with the 21$\mu$m upper limit.  If it is less than 400 mag, one
has to invoke the above mentioned background cloud to explain the sub-mm fluxes.
These model results are also consistent with our fits to the ammonia lines
described in \S\ref{subsec:amm}

It is important to note that the model spectrum for the FIR does not rise again
at shorter wavelengths and cannot account for the 8.2$\mu$m MSX and the Spitzer
fluxes.  Here an independent source is required.  We propose that besides the
star, there is a second blackbody emitter close to it.  Its temperature is
approximately $T= 550$ K, its total luminosity 150 L$_\odot$, and it suffers 150
mag of visual extinction.  This is below the minimum estimate for $A_{\rm V}$ of
the envelope, but the latter may, of course, be patchy.  It is natural to
associate the second blackbody with an accretion disk of about 12 AU radius
(outer).
Considering the paucity of the data, more refined calculations seem, at present,
unwarranted.

The dashed line that fits the long wavelength data in Fig.\ref{fig:sed} is a modified
Planck curve of the form $\nu^2B_\nu(T)$ with $T= 19$~K, possibly the emission
of the background cloud with a total mass of 350~$ M_\odot$.

\label{fig:5}

\subsubsection{2MASS sources\label{subsec:2mass}}

As shown in Fig.\ref{fig:g11image}, the \meth\ peak is offset from the 2MASS source by $\sim
2''$.  It is highly unlikely that the ATCA observations have errors on the
order of 2$''$ (cf \S 3.1).  Though the rms uncertainty in the case of 2MASS
data is on the order of 0.4$''$, we cannot completely rule out that the embedded
object triggering maser emission is not associated with the 2MASS source.
Because of the high column density towards the centre of the core, it could be
that we are seeing light from the embedded object scattered out from the edges
of the NIR core.  We now discuss such a situation in detail.

The 2MASS J and H bands reveal 3 faint point sources detached from the sub-mm
peak, while at K, where the resolution is inferior, there is only one single
source.  These 2MASS points lie far above any possible fits to a spherical cloud
model.  We first investigated whether they can come from an unrelated foreground
star of lower bolometric luminosity but suffering less obscuration.  Analysing
the observed JHK fluxes in a colour-colour diagram, we found that they cannot be
explained by reddening, so we discarded the possibility of a foreground star.

Instead, we propose that we are dealing with three knots of scattered light that
escapes from the star into an optically thin cone above a hypothetical disk.  We
estimate the flux scattered by the cone at frequency $\nu$ from the formula

\begin{equation}
S_\nu  = {f_\nu L_\nu \over 4 \pi D^2},
\end{equation}

where $D$ is the source distance and $L_\nu$ the stellar spectral luminosity, so $L =
\int L_\nu d\nu$.  If $\Omega_{\rm cone}$ denotes the solid angle of the cone
and $\tau_\nu^{\rm sca}$ its (small) scattering optical depth, then $f_\nu=
\tau_\nu^{\rm sca}\Omega_{\rm cone}/4\pi$.  Because of the basically unknown
geometry, our estimates are very rough.  But assuming the star to be a
blackbody, only about one percent of the light has to be scattered (the exact
number depends on the stellar temperature).

\begin{figure*}[h]
\centering \includegraphics[width=\linewidth,angle=0]{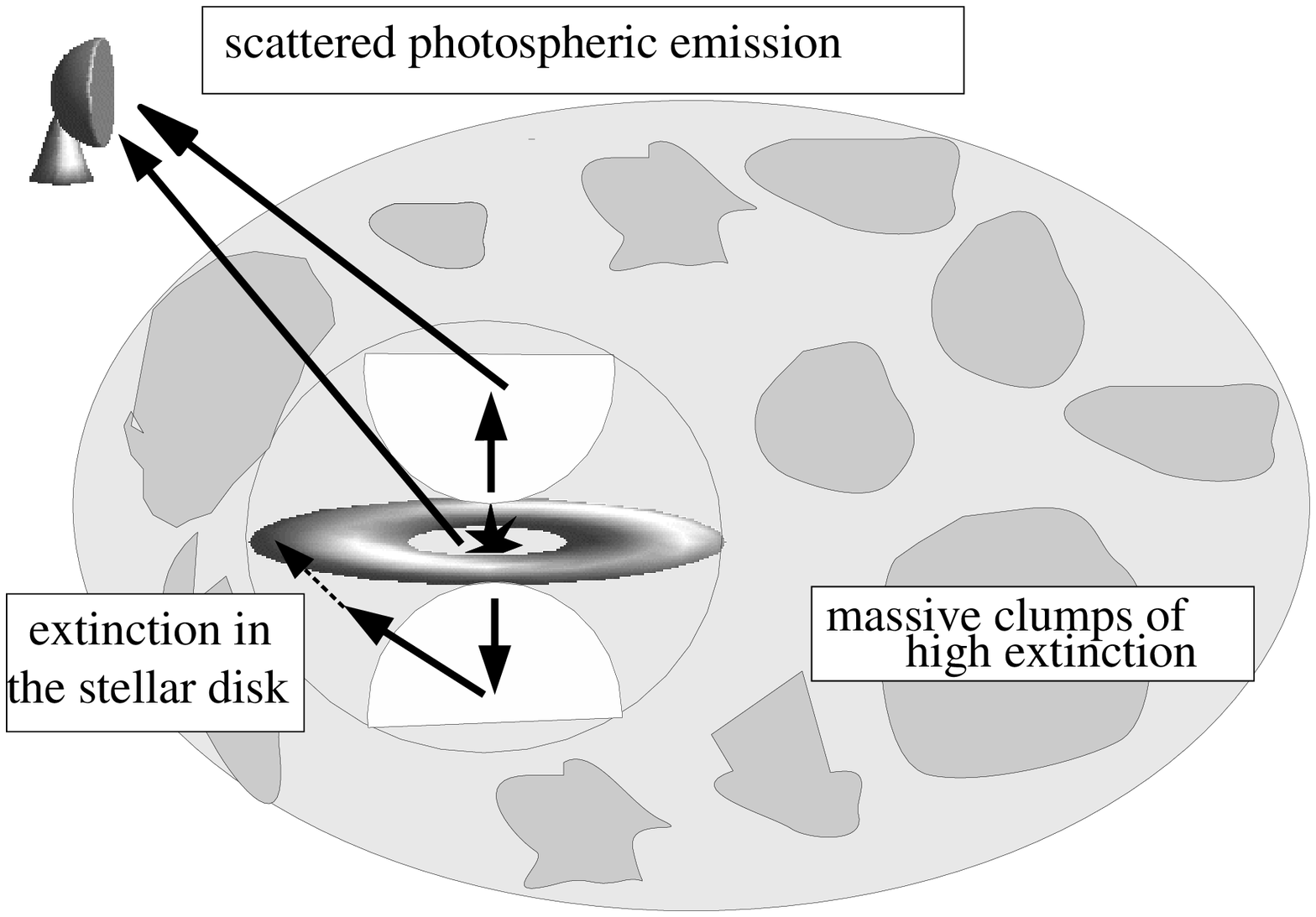}
  \caption{Illustration of an intuitive model of the source geometry
needed to explain the NIR emission. The protostar is part of a massive
(several 100~M$\odot$) clumpy cloud complex. The optically thin
outflow cone scatters the stellar light at NIR wavelengths toward
the observer. The stellar
photons scattered from the cone below the disk is extincted by the optically 
thick disk; consequently, we observe a uni-polar nebula at NIR.}
\label{fig:geometry}
\end{figure*}

As regards the wavelength dependence, scattering of pure stellar light
would suggest decreasing fluxes at longer (K) wavelengths.  However,
foreground extinction and, more importantly, hot dust from an
accretion disk ($T \sim 1000$\,K) could easily explain the observed
spectral shape at near IR wavelengths. In view of the geometrical
requirements posed by the near IR fluxes, we propose a more realistic
configuration.  In such a model, the protostar is part of a massive
(several 100~M$\odot$) clumpy cloud complex.  Such a patchy dust
distribution is shown in Fig.\ref{fig:geometry}. As seen in
Fig.\ref{fig:g11image}, the emission at NIR is concentrated at only
one edge of the BIMA 3mm emission. In the configuration envisaged
in Fig.\ref{fig:geometry}, the stellar radiation is scattered by the
dust clump above the disk at NIR wavelengths toward the
observer. However, the stellar photons scattered from regions below
the disk suffer severe extinction from the disk and fail to reach the
observer at NIR wavelengths, resulting in a uni-polar nebula in NIR.

\label{fig:6}
 
 We therefore regard the JHK knots as further indirect evidence of
a stellar disk.

\section{Conclusion\label{sec:conclusion}}
Our main result is the detection of what is most likely a high-mass
young stellar object within the IRDC G11.11. If the luminosity we
derived is an overestimation that corresponds to an intermediate star,
this would be the first reported case of CIIMM detection towards an IM
mass-star-forming region. This evidence comes from the
following observations and arguments:

\begin{itemize} 

\item 
\meth\ and \WAT\ masers appear during the earliest phases of massive star
formation, so their detection in G11.11 is evidence of an embedded
object. 

\item 
The systematic velocity gradient and the linear feature in the P-V diagram of
the \meth\ lines is a signature of a disk encompassing a protostar.

\item 
The higher excited \AMM\ (3,3) thermal line points towards a compact
and hot central component.  The thermal \meth\ emission (Leurini et
al. in preparation) suggests an outflow, in agreement with the claim
by \citet{johnstone2003:g11}, based on H$_2$CO obervations.  However,
\citet{johnstone2003:g11} do not find any evidence of high-mass star
formation.

\item 
The strong FIR fluxes ($\ge 60$ Jy) derived from analysing the high resolution
IRAS data imply a powerful central source ($> 1000$ L$_\odot$). 

\item
The radiative transfer additionally requires blackbody emission of
about 550 K from a separate, less embedded source, probably from a
disk, in order to reproduce the Spitzer data between 3.6 and
8.0$\mu$m.  A non--spherical geometry is also indicated by the 2MASS
NIR fluxes, which can only be explained by scattering and not through
reddening.

\item
The high-extinction estimates also suggest an extremely young object, probably
the Class 0 equivalent of a high--mass protostar.
\end{itemize}

\appendix

\section{Model}

We iteratively solved the energy equation

$$ \int \kappa_\nu J_\nu \, d\nu = \int \kappa_\nu B_\nu(T_{\rm d}) \,
d\nu $$, together with two integral radiative transfer equations

$$ I^+(\tau) \ = \ e^{-\tau} \left( I^+(0) \ + \ \int_0^\tau S(x)
\,e^x\, dx \right) $$
$$ I^-(t)\ =\ e^{-t} \ \int_0^t S(x) \,e^x \, dx $$.

\noindent Here $\kappa_\nu$ denotes the dust extinction coefficient, $J_\nu$ the mean
intensity, and $B_\nu(T_{\rm d})$ the Planck function at dust temperature $T_{\rm
d}$.  In the formulae for the radiative transfer, $I^+(\tau)$ is the intensity
directed towards the observer at optical depth $\tau$, frequency $\nu$, and
impact parameter $p$, while $I^-$ refers to beams away from the observer.  The
source function $S(\tau)$ includes a scattering term.  The optical thickness
$\tau$ and geometrical length $z$ are related through $d\tau = \rho \kappa \,
dz$, where $\rho$ is the density. Index $\nu$ has been dropped for
convenience of writing.

There are two boundary conditions.  One states that the
standard interstellar radiation field (ISRF) impinges at the cloud surface ($I^-_\nu= I^{\rm
ISRF}_\nu$), the other reads $I^+ -I^-=0$ at $\tau=0$ (where also $z=0$) at
impact parameters $p$ greater than the stellar radius $R_*$, and that $I^+ -I^- =
F_*\nu/\pi$ at the stellar surface ($F_{*\nu}$ is the stellar flux at frequency
$\nu$).

\acknowledgements{ We thank the ATNF for the observing time allocation at the ATCA.  The
Australia Telescope is funded by the Commonwealth of Australia for operation as
a National Facility managed by CSIRO.  We are grateful to M.Voronkov for making
our ATCA observations possible.  This research has made use of the NASA/ IPAC
Infrared Science Archive, which is operated by the Jet Propulsion Laboratory,
California Institute of Technology, under contract with the National Aeronautics
and Space Administration.  T.Pillai was supported for this research through a
stipend from the International Max Planck Research School (IMPRS) for Radio and
Infrared Astronomy at the University of Bonn.  T.Pillai thanks her colleague
J.Kauffmann for all the helpful discussions.}

\bibliographystyle{aa}
\bibliography{bib_astro}
\end{document}